\documentstyle[preprint,aps]{revtex}
\newcommand{\cb}{\langle  |c|\rangle}
\newcommand{\cl}{\langle \ln  |c|\rangle}
\newcommand{\gl}{\langle g_L\rangle}
\newcommand{\gt}{\langle \tilde{g}_{L}\rangle}

\newcommand{\cbt}{\langle |\tilde{c}_\alpha|\rangle }
\newcommand{\tr}{{\rm tr}}
\begin{document}
\draft

\title{Level Curvatures and Conductances: A Numerical
Study of the Thouless Relation}
\author{D.~Braun$^{(1)}$, E.~Hofstetter$^{(2)}$, G.~Montambaux$^{(3)}$,
and A.~MacKinnon$^{(2)}$ }
\address{$^{(1)}$ FB7, Universit\"at--GHS Essen, 45\,117 Essen, Germany\\
$^{(2)}$ Blackett Laboratory, Imperial College, London SW7 2BZ, England\\
$^{(3)}$ Laboratoire de Physique de Solides, associ\'e au CNRS, Universit\'e
Paris Sud, 91\,405 
Orsay, France}
\maketitle
\thispagestyle{empty}
\bigskip

The Thouless conjecture states that the average
conductance of a disordered metallic
sample in the diffusive regime can be related to the sensitivity of the
sample's 
spectrum to a change in the boundary conditions. 

Here we present results of a direct numerical study of the
conjecture for the Anderson model.  They were obtained
by  calculating  the Landauer--B\"uttiker conductance  $g_L$ for a sample
connected to perfect leads and  the
distribution of level curvatures for the same sample in an isolated ring
geometry, when the ring  is pierced by an Aharonov--Bohm flux.  In the
diffusive regime ($L\gg l_e$) the
average  conductance $\gl$  is proportional to the
mean absolute curvature $\langle |c| \rangle$: $ \gl  = \pi
\langle| c |
\rangle / \Delta$, provided the system size $L$ is large enough, so that the
contact resistance can be neglected. $l_e$ is the elastic mean free path, 
$\Delta$ is the
mean level spacing. When approaching the ballistic regime,
 the limitation of the conductance due to the contact resistance
becomes essential and expresses itself in a deviation from the above
proportionality. However, in both regimes and for all system sizes the same
proportionality is recovered when
the contact resistance is
subtracted from the inverse conductance, showing that the {\em curvatures
measure 
the conductance in the bulk}.  
In the localized regime, the mean
logarithm of the absolute curvature and the mean logarithm of the
Landauer--B\"uttiker conductance are proportional.

\newpage

\section{Introduction}\label{intro}
It has been shown by Thouless and Edwards in the  70's that the conductance of
a disordered diffusive system can be related to the dependence of
the energy levels to a change in the boundary
conditions\cite{edwards72,thouless74}.
A physical realization of this change in the boundary conditions is made
when the sample is closed to a ring and pierced by a
Aharonov--Bohm flux $\phi$. In such a case the wave function must obey
the condition  $\psi(x+L)=\psi(x)e^{i\eta}$ where $\eta=2 \pi \phi/\phi_0$.
$\phi_{0}=h/e$ is the flux quantum.
Thouless found that the average conductance is
proportional to the width of the
distribution of the curvatures of the energy levels when $\eta$ is the
perturbation parameter.
This relation is based on the similar structure of the Kubo expression for
the conductance and of the curvature of energy
 levels when the boundary conditions are changed.
In the diffusive regime, the average absolute curvature of the energy levels
is proportional to the diffusion 
coefficient and thus  to the conductance.
Although this conjecture was derived under oversimplified
assumptions, its basic idea  has proven to be very powerful and has
become a keystone in our understanding of localization.\\

Meanwhile, other measures of the conductance in terms of the response of the
system's energy spectrum to a change of the boundary conditions have been 
derived.
These spectral measures of the conductance  were all obtained for the
diffusive regime. They all measure  the diffusion 
constant in the bulk, and are in this sense 
equivalent to each other. Indeed, it has been shown that the 
Kubo conductance can be expressed as the average square of the first
derivative of the energy levels with respect to flux
\cite{wilkinson88,akkermans92,simons95}. The
average has to be 
taken over both flux and disorder. The relation between these quantities
and the curvature distribution  was  studied both 
numerically and analytically \cite{akkermans92,us94,fyodorov95}.\\

However, so far very little effort has been devoted to comparing spectral
measures of the conductance to the  customary conductance formulas based on
transport 
considerations, like in particular the Landauer--B\"uttiker formula.
Exceptions are Ref.\cite{zharekeshev??}, where small 1D
systems were treated numerically, and Ref.\cite{casati96} devoted to the study
of random banded matrices. This  is  surprising, since
spectral measures of the conductance have been widely used in the mesoscopic
community. Yet, it is well known that the conductance of a mesoscopic sample
depends sensitively on the measurement geometry and on the way the leads
are attached to the sample.
 On the other hand, the spectral
measures of conductance are completely insensitive in this respect, since
the sample is closed to a ring and no leads are attached. The system is then
in fact a different one. Its spectrum is discrete when the system is
finite, whereas the system attached to the leads always has a continuous
spectrum. \\

How can  it then be possible that the width of the curvature distribution
measures the conductance obtained in a transport measurement? 
In this paper we find that the appropriate  conductance
is obtained in a situation where the system is  connected to leads with
the same transverse width and with maximal transmission coefficient. We
refer to this situation as "maximal coupling". It is analogous to the
``matching wire'' condition defined by Economou and Soukoulis
\cite{economou81} and also used in ref.~\cite{casati96}.
In this case  the
mean absolute curvature is proportional to the Landauer--B\"uttiker
conductance. This proportionality  holds
if the system is large enough and the disorder strong enough such that the
contact resistance can be 
neglected.
In the ballistic regime, the contact resistance is always 
important and destroys the proportionality. However, after subtracting the
contact resistance from the total inverse conductance, the remaining bulk
conductance 
is proportional to the mean absolute curvature. The proportionality
coefficient 
is the same in the 
diffusive and in the ballistic regime for all system sizes, as long as the
samples have similar geometry. This shows that
{\em the curvatures measure a bulk conductance}. In the localized regime, it
is the mean logarithm of the absolute curvatures and the conductances that
are proportional to each other.\\
  
In the next section we briefly review the 
different  definitions  of 
the conductance.  In section \ref{num} we  discuss our numerical method and 
in section \ref{secResults} we present the results of
extensive numerical simulations, in which we calculated the level curvatures
from a perturbative formula and the conductances with the
Landauer--B\"uttiker formula. The 
conductances extend over seven orders of magnitude and allow us to study the
diffusive as well as the ballistic and localized regimes. We conclude in
section \ref{concl}.

 \section{Kubo, Landauer and Thouless conductances} \label{RLT}
\subsection{Kubo and Thouless conductances}

Let us first recall the main line of the Thouless derivation.
On the one hand, the d.c. Kubo conductivity can be
written as\cite{kubo}:

\begin{equation} \label{sigmaK}
 \sigma =  {\pi   e^2 \hbar \over m^2 V} \sum_{\alpha,\beta}
 |p_{\alpha\beta}|^2 \delta (E_F - \varepsilon_\alpha)  \delta (E_F -
\varepsilon_\beta) \ \, \end{equation}
 where ${\varepsilon_\alpha}$ are single energy levels
 and $p_{\alpha\beta} = \langle \alpha | \hat{p}| \beta \rangle$ are the
matrix element of the   momentum operator.  $V = L^d$ is the volume and
$E_F$ the Fermi energy.
Strictly speaking, this expression is zero for a finite system
\cite{imry86}. To get a 
finite $\sigma$, the $\delta$ functions must have a finite width larger than
the inter-level spacing.
Under this condition and
assuming that the matrix elements $p_{\alpha\beta}$ are decorrelated
from the $\varepsilon_\alpha$ \cite{RMT}, the average conductivity
is given by: \begin{equation} \label{Gk1}
\langle  G_K \rangle = \langle \sigma\rangle L^{d-2}  =  {\pi   e^2 \hbar
L^{2d-2} \over m^2}   \langle |p_{\alpha\beta}|^2  \rangle
\rho_0^2\,.
\end{equation}
${\langle}$...${\rangle}$ represents an average  over the
disorder.
$\rho_0$ is the average density of states per unit volume at the Fermi energy.
The dimensionless conductance $\langle g_K \rangle$ can be written
as\cite{remark1}:
\begin{equation} \label{Gk2}
\langle g_K \rangle \equiv {\langle G_K \rangle \over e^2 /h}=2 \pi {E_c
\over \Delta} \end{equation}
where $\Delta=1/(\rho_0 L^d)$ is the mean level spacing, and the
Thouless energy $E_c$ is given by $E_c = \hbar D / L^2$, $D$ being the
diffusion coefficient \cite{thouless74}. The second equality in formula
(\ref{Gk2}) is nothing but the Einstein relation $\sigma=e^2D\rho_0$. All
these  
quantities are defined for a given Fermi
energy and will in general depend on $E_F$.

 On the other hand, under the change in the boundary conditions
$\psi(x+L)=\psi(x)e^{i\eta}$, the curvature of a given
 energy level $\varepsilon_\alpha$   at
the origin ($\eta = 0$) is given exactly by  perturbation
expansion in  $\eta$:
\begin{equation} \label{cn}
c_\alpha=\left ( {\partial^2 \varepsilon_\alpha\over
\partial \eta^2}\right )_{\eta=0}
 ={ \hbar^2\over m L^2}
+{{{ 2\hbar}^2}\over{{m^2{L^2}}}} \sum_{\beta \neq \alpha}
{ |p_{\alpha\beta}|^2
\over \varepsilon_\alpha-\varepsilon_\beta}\,.
\end{equation}
In order to relate the width of   the curvature distribution to the
diffusion coefficient,
Thouless assumed first that the  energy levels
$\varepsilon_\alpha$ are not correlated with the matrix elements
$p_{\alpha\beta}$. Replacing then $|p_{\alpha\beta}|^2$ by its average
value, the distribution of the curvatures is that of
$1/(\varepsilon_\alpha-\varepsilon_\beta)$. Secondly, assuming    that the
energy levels themselves are
not correlated, the sum in
eq.(\ref{cn}) gives rise to a Levy law for the 
distribution of the
curvatures   
in the limit of infinitely many levels \cite{edwards72}. 
It  has the Cauchy form $P(c) = (\gamma_0 / \pi)/(\gamma_0^2+c^2)$
with a width $\gamma_0$ given by
\begin{equation} \label{curvature}
\gamma_0 = {2 \pi \hbar ^2  \over m^2 L^2 }
\frac{\langle {|p_{\alpha\beta}|^2\rangle}}{ \Delta}\,.
 \end{equation}
 Comparison between the equations
 (\ref{Gk1}) and (\ref{curvature}) gives the relation between the
dimensionless average conductance $\langle g_K \rangle = \langle G_K
\rangle h / e^2$  and the width of the distribution of
curvatures, known as the Thouless relation\cite{thouless74}:
$\langle g_K\rangle  = \pi {\gamma_0 \over \Delta}$.
However, it is now
known that the energy levels are strongly correlated in a
metal so that
the curvature distribution does not have the Cauchy form.
Instead, it is given by: 
\begin{equation}\label{curvatured}
P_\beta(c) = {N_\beta \over (\gamma_\beta^2 +
c^2)^{(\beta+2)/2} } 
\end{equation} 
Here, $\beta=1$ if there is
time-reversal symmetry and $\beta=2$ if
time-reversal symmetry is broken.
$N_\beta$ is a normalization coefficient. This form
has first been guessed by Zakrzewski and Delande\cite{zakrzewski93}
to
fit numerical calculations on various models exhibiting chaotic spectra.
 It has
been proven analytically by von Oppen\cite{vonoppen94} for random
matrices of the form $H(\lambda) = H + \lambda K$ where  $H$ and $K$ are
random matrices belonging to the same symmetry class ($\beta=1$ for the
Gaussian 
Orthogonal Ensemble, GOE;  $\beta=2$ for the Gaussian
Unitary Ensemble, GUE).
$\lambda$ is the perturbation parameter. Recent numerical calculations have
shown that this distribution is also characteristic of metallic
spectra when the perturbation parameter is an AB  flux
$\phi$\cite{us94}.
In particular, in the limit where $\phi \rightarrow 0$, the
distribution is still the GOE distribution ($\beta=1$ in eq.
(\ref{curvatured}))\cite{us94}.
This has been proven analytically by Fyodorov and Sommers
who also found that
there are no corrections of order $\Delta/E_c$\cite{fyodorov95}.\\

The normalized distribution in zero field is thus:
\begin{equation}\label{curvature1}
P_1(c) = {1 \over 2}{\gamma_1^2 \over (\gamma_1^2 +
c^2)^{3/2} } \end{equation}
Fyodorov and Sommers have shown that the width of this distribution can
be related to the diffusion coefficient \cite{fyodorov95}. They find that
the width $\gamma_1$ of this 
distribution for a three dimensional ring 
is given by $\gamma_1 = 2 E_c$. Using now  eq.~(\ref{Gk2}) relating $E_c$ to
the Kubo conductance, one deduces
$\gamma_1=\Delta \langle g_K \rangle
/\pi$. To characterize this width it is convenient to introduce the average
absolute curvature $\langle|c| 
\rangle=\gamma_1$, so that 
{\em the Thouless conductance defined as} 
\begin{equation}
\label{g1}   
\langle g_T\rangle  \equiv  \pi {\langle |c|\rangle \over \Delta}  
\end{equation}
{\em equals the Kubo conductance}: $\langle g_T\rangle = \langle g_K \rangle$.\\
It has to be noted: first, all conductances considered so
far are average  conductances, the average extending over disorder
realizations. 
Secondly, the equation $\langle g_T\rangle = \langle g_K \rangle$
holds so far only for the diffusive regime,
since $\gamma_1=2E_c$ was derived in \cite{fyodorov95} for the diffusive
regime. We will see that the equation has to be modified in the ballistic
and localized regimes.

\subsection{Landauer-B\"uttiker conductances}

Another way to express the conductance has been introduced by Landauer.
He related this quantity to the scattering properties of the disordered system,
when it is {\it connected} to {\it incoherent}
reservoirs through {\it ideal leads}. This approach ideally suits transport
through finite mesoscopic systems and shows the importance of the
measurement geometry.
For one dimension, Landauer derived the dimensionless
conductance $\tilde{g}_L$ \cite{landauer}:
\begin{equation} \label{Gl1}
\tilde{g}_L \equiv {\tilde{G}_L \over e^2/h}=
 {T \over 1-T}\,.
\end{equation}
This conductance is the
ratio
$\tilde{G}_L=I/(\mu_A- \mu_B)$ where
$\mu_A$ and $\mu_B$ are the chemical potentials of ideal leads
 attached to the barrier. $T$ is the transmission coefficient through  the
disordered region.  $\tilde{g}_L$ diverges for an ideal, clean sample.
On the other hand, Economou and Soukoulis, trying to derive this formula
from linear response 
theory (Kubo formula), found \cite{economou81}:
\begin{equation} \label{geco}
g_L= {G_L \over e^2/h}=T
\end{equation}
instead. In this case,
$G_L=I/(\mu_1-\mu_2)$,
where $\mu_1$ and $\mu_2$ are the chemical potentials of the reservoirs
\cite{imry86}. 
Eq.(\ref{Gl1}) describes a four-terminal measurement in one
dimension, that is a measurement with separate current and voltage probes
\cite{anderson81}. 
$g_L=T$ describes a
two-probe measurement, where only two leads are attached to the sample and
serve as current and voltage probes at the same time. The 
remaining finite resistance at zero disorder ($g_L=1$) is a ``contact
resistance'' 
which has its origin in the coupling of the  sample  to the incoherent
reservoirs \cite{imry86,buttiker85}. This resistance cannot be avoided in a
two-probe measurement. 
One may therefore think of the total resistance $G_L^{-1}$ in such a 1D
two-probe 
geometry as being the sum of the contact resistance $h/e^2$ and a ``bulk
resistance'' $\tilde{G}_L^{-1}$. The latter vanishes when the disorder goes
to zero and is 
identical with the original Landauer contribution:
\begin{equation}\label{contact1} 
G_{L}^{-1} = \tilde{G}_L^{-1} + {h \over e^2}
\end{equation}
Fisher and Lee generalized eq.(\ref{geco}) to the multi-channel case
\cite{fisher81}:  
\begin{equation} \label{g3}
g_L=g_K=
  \sum_{i=1}^M T_i=\tr tt^+\,.
\end{equation}
$T_i$ is the total transmission probability in the $i^{th}$
channel, $t$ the transmission matrix and $M$  the number of channels.
When the disorder in the sample goes to zero, $g_L$ is limited by the
number of open channels.
Today there is  a general agreement that eq.(\ref{g3}) describes a two-probe
measurement in a multi-channel geometry. In our 
numerical simulations we will focus on this situation and use eq.(\ref{g3}) 
for the numerical evaluation of the conductance.\\ 
Comparing equations (\ref{g1}) and (\ref{g3}), one gets
a relation between the average Landauer--B\"uttiker conductance and  the
width of the 
curvature distribution: \begin{equation} \label{g4}
\gl  =
  \pi {\langle |c | \rangle \over \Delta}\,.
\end{equation}

In the two-probe multi-channel geometry we will consider in the following
one might again decompose the total resistance into a sum of a
contact
resistance plus a bulk  resistance, the latter being entirely  due to the
motion in the bulk of the sample.  In straight generalization of
eq.(\ref{contact1}), 
it is then natural to define the bulk
conductance  $\tilde{G}_L=\frac{e^2}{h}\tilde{g}_L$ by
\begin{equation}
G_{L}^{-1}= \tilde{G}_L^{-1} + R_c\,,
 \end{equation}
where $R_c = h /(M e^2) $ is the ``contact resistance'' for the
multi-channel system \cite{imry86,buttiker88}. We have then 
\begin{equation}\label{gtilde}
\tilde{g}_L=
 { \sum T_i  \over 1 - {\sum T_i\over M}}\,.
\end{equation}
In the diffusive regime, the effective number 
of conducting
channels, $M_{eff}=\sum T_i$, is much smaller than $M$: $M_{eff}= M l_e / L$
where $l_e$ is the elastic mean free path\cite{imry86b}.
 Consequently, $g_L$ and $\tilde{g}_L$ are almost identical in the
 diffusive regime, the relative deviations being of order $l_e/L$.
 However, in the ballistic regime they behave very differently:  
$\tilde{g}_L \rightarrow \infty$ and $g_L \rightarrow M$ in the limit of
zero disorder.\\

For more than one channel $\tilde{g}_L$ has not the simple and general
interpretation 
of the conductance measured in a four-probe measurement. Indeed,
that in the multi-channel case not only  the number  of leads but also the way
(e.g.~under what angles) they are attached influences the measured
conductance, such that a general four-probe formula might not even exist
\cite{buttikerpc2}. Similarly $\cb$
cannot correspond to any particular four-probe conductances, since
it is  an intrinsic property of the disordered region.
 We will show
that for a finite system $\cb$ is proportional to
$\gt$: 
\begin{equation} \label{rightp}
\gt=\pi\frac{\langle |c|\rangle}{\Delta}\,.
\end{equation}

\section{The numerical method}\label{num}
The starting point of our analysis is the Anderson tight--binding Hamiltonian
$H$ of a
disordered system on a square lattice of $L_{x}\times L_{y}\times L_{z}$
sites. 
 For the curvature calculation the system is closed to a
ring and pierced by an Aharonov--Bohm flux $\phi$:
\begin{equation} \label{HAnd1}
H=\sum_i e_i |i\rangle\langle  i|+u \sum_{<ij>} |i\rangle\langle
j|
+
u\sum_{\stackrel{<ij>}{i_{x}=L,j_{x}=1}} (
e^{i\eta}|i\rangle\langle j|+h.c.)\,. \end{equation}
The $e_{i}$ are distributed uniformly and independently in an interval between
$-w/2$ and $w/2$. $<ij>$ denote next nearest neighbors,  $u$ is the
hopping matrix element which we set equal to one in the following, and $w$
is the disorder parameter.  The last
sum in eq.(\ref{HAnd1}) is over the set of sites on the two boundaries
limiting the open 
sample in $x$--direction. Hopping between these boundary sites arises when the
system is closed to a ring and includes a phase factor 
$e^{i\eta}$. For $\phi=0$ or entire multiples of the flux quantum, one recovers
periodic boundary conditions.\\ 
For the calculation of $g_L$, the system is open and coupled to perfect
leads. The last sum in 
eq.(\ref{HAnd1}) is then missing. This 
 is the only difference
between the two Hamiltonians. In particular, for the numerical
implementation the same random number generator was used for the diagonal
matrix  
elements in both situations. 

\subsection{Curvatures}
In the diffusive regime, the curvatures can be evaluated by replacing
differentials by small flux 
differences whose values are varied for control in a suitable way \cite{us94}.
This procedure has the  numerical advantage that only eigenvalues, not
the eigenvectors are needed. However,  it is very difficult
to control in the ballistic and  in the localized regime. We adopted
therefore a 
routine based on an exact perturbative formula corresponding to
eq.(\ref{cn}). In fact, treating 
$\eta$
in eq.~(\ref{HAnd1}) as a perturbation up to second order, one  finds for the
curvatures at zero flux 
\begin{equation}
  \label{PT}
{c_\alpha \over 2} =\sum_{\stackrel{<ij>}{i_{x}=L_{x},j_{x}=1}}\langle
\epsilon_\alpha|i\rangle \langle  j|\epsilon_\alpha\rangle +\sum_{\beta\ne
\alpha}\frac{1}{\epsilon_\beta-\epsilon_\alpha}\left(\sum_{\stackrel{<ij>}{i_{x}
=L_{x},j_{x}=1}}\langle
\epsilon_\beta|i\rangle\langle  j|\epsilon_\alpha\rangle-\langle  
\epsilon_\beta|j\rangle\langle 
i|\epsilon_\alpha\rangle\right)^2  \,,
\end{equation}
where $\epsilon_\alpha$ and $|\epsilon_\alpha\rangle$ denote the eigenvalues and 
eigenvectors of the
Hamiltonian at zero flux, respectively. Higher order terms vanish since
$\eta=0$. In the two directions perpendicular
to the transport direction, periodic boundary conditions were used. 
Formula (\ref{PT}) is exact as long as
$\epsilon_\beta\ne \epsilon_\alpha$. Thus, for a finite system, where level 
repulsion is always
present at sufficiently small energy scales \cite{shklovskii93},
(\ref{PT}) remains valid also in the localized and the ballistic regime.
Besides rounding errors which can be neglected here the only remaining
errors in the calculation of $\langle |c_\alpha|\rangle$ are statistical errors 
that
can be controlled by increasing the number of disorder realizations. We used
up to 1000 disorder realizations for system sizes of $6\times
6\times 6$ sites and still about hundred for $10\times 10\times 10$ sites.
Relatively, the remaining statistical errors in the diffusive and ballistic
regimes were  of the order of $10^{-2}$, which we checked by varying the number 
of
disorder realizations. As
eq.(\ref{PT}) indicates, {\em all} eigenvalues and eigenvectors are needed
for the calculation of a single curvature. Realizations where our
Lanczos routine failed to find all eigenvalues and eigenvectors were
therefore discarded. 

\subsection{Conductances}
The conductance $g_L$ was calculated  from eq.(\ref{g3}) by the Green's
function recursion technique \cite{macKinnon85}. The Green's function
connecting  the 2
ends of a strip can be calculated recursively using the equations
\begin{eqnarray}
{\bf G}_{N,N}^{(N)} &=& \left[Z - {\bf H}_{N} 
- {\bf u}^\dagger{\bf G}_{N-1,N-1}^{(N-1)}{\bf u}\right]^{-1}\\
{\bf G}_{1,N}^{(N)} &=& {\bf G}_{1,N-1}^{(N-1)}{\bf u}{\bf
G}_{N,N}^{(N)}
\end{eqnarray}
where ${\bf G}_{N,N}^{(N)}$ represents the sub-matrix of the Green's
function between sites on the $N$th slice of a strip of length $N$,
${\bf G}_{1,N}^{(N)}$ is the corresponding sub-matrix between
sites on the $1$st and $N$th slices, and ${\bf H}_{N}$ represents the
Hamiltonian of the $N$th slice alone.  The system can be embedded in
semi-infinite leads by choosing the initial values of the 2 Green's
functions to represent the end of a semi-infinite wire and by adding a
final slice for which the Hamiltonian of the slice is replaced by the
self-energy matrix for another semi-infinite wire. Having the Green's 
functions one can derive the transmission matrix $t$ \cite{fisher81}
and then the conductance $g_{L}$.

\section{Results}\label{secResults}
\subsection{Energy dependence}
Without averaging over energy, 
both $\langle g_T\rangle$ and $ \gl$ are energy dependent:
$\langle g_T(E)\rangle$ and $\langle g_L(E)\rangle$. The variation of
$\langle g_L(E)\rangle$ is smooth and is due to the energy dependence of the
DOS and of the number of channels $M$.
 The
energy dependence of $\langle g_T(E)\rangle$ arises from the variation with
energy  of both
$\langle |c|\rangle$ and $\Delta (E)$, where the latter quantity  is the mean
level spacing at a 
given energy (averaged over disorder only). 
In order
to get the conductance at a given energy, we therefore rescaled the
curvatures with an energy dependent $\Delta$: 
$\tilde{c}_\alpha = c_\alpha/\Delta(\epsilon_\alpha)$. The disorder averaged 
DOS, $1/\Delta(E)$ , was obtained by
the standard method of fitting the spectral staircase (integrated DOS) to a
polynomial.\\ 

After averaging over 1000 disorder realizations (in the case of systems with
$6\times 6\times 6$ sites), the fluctuations of  $\cbt= \langle
|c_\alpha|\rangle / \Delta(\epsilon_\alpha)$ as a function of energy  
turned out to be still much more pronounced than those of $\langle
g_L(E)\rangle $.  
This is not too surprising, as it is well known that in the diffusive regime 
the conductance distribution (which is a universal Gaussian distribution
\cite{stone85,altshuler85} with a width of the order of the conductance
quantum)  and the curvature distribution  (see section \ref{intro}) are very
different.  
Thus, when using just one disorder realization, the fluctuations of the
function $|c_\alpha(\epsilon_\alpha)|$ will be much larger than those of
$\langle g(E)\rangle$, due to
the long $1/c_\alpha^3$ tails of the curvature distribution. For a
finite number of realizations this difference will still persist, and only
when  averaging over infinitely many disorder realizations  the
energy dependence of  $
\gl$ 
should follow that of $\langle |\tilde{c}_\alpha|\rangle$.   
Having in mind that even 1000 disorder 
realizations did not suffice to reduce the fluctuations of $\langle
g_T(E)\rangle $ to a
level comparable to those of $\langle g_L(E)\rangle$, it seems very difficult to 
check
the Thouless conjecture  
in the stronger sense for a given energy with the current computing power 
available. We therefore  averaged $g_L$ and $g_T$ not 
only over the realizations but over a band of energy 
$\Delta E$ comprising typically about the central half of the spectrum as
well. We checked that increasing the size of the system 
or the number of realizations allow to decrease $\Delta E$ to obtain the same
results. This suggests that our results are independent of $\Delta E$.
In the following, $\langle \ldots \rangle$ will stand for the combined
disorder and  
energy average. Care was taken in order to average both curvatures and 
conductances over exactly the same energy interval.

\subsection{Curvature distribution} 
In the diffusive regime, the distribution of the curvatures is well
described by eq.(\ref{curvature1}). Thus,  $\cb$ is a good
measure of the width of the curvature distribution. \\
Outside the diffusive
regime the curvature distribution was not known so far,  and one might
wonder whether $\cb$ is still well defined. We therefore calculated $P(c)$
numerically for both the ballistic and the localized regimes.
 Fig.\ref{pofcbal} shows $P(c)$ 
for a system in the ballistic regime ($6\times 6\times 6$ sites, $w=1.0$,
4000 disorder realizations) and the prediction of  eq.(\ref{curvature1}),
where $\gamma_1$ was determined  as $\gamma_1=\cb$ (no fitting parameter). 
Eq.(\ref{curvature1}) works 
well for large curvatures and shows that in the ballistic regime $P(c)$ has
$1/c^3$ tails as in the diffusive regime. For small curvatures deviations
from eq.(\ref{curvature1}) in the form of non--universal features appear and
the distribution develops two maxima. 
A relative minimum appears at zero curvature. These deviations become
even more pronounced  for smaller disorder. Altogether we conclude that
$\cb$ can still serve as a measure 
for the width of the curvature distribution, even in the ballistic regime. 
 \\

In the localized regime at least two different numerical works favor a
log--normal curvature distribution 
\cite{karol94,canali96}. Analytical evidence for a  log--normal
distribution at least 
for small curvatures in 1D is given in \cite{yan96}.
On the other hand, one might suspect that
Thouless' original result of a Cauchy distribution due to uncorrelated
eigenvalues might apply to the localized regime. Such a distribution would
of course spoil the use of $\cb$ as a measure of the Thouless conductance. 
We therefore reexamined this question numerically. As shown in
Fig.\ref{pofclocLor}, a Cauchy distribution can be ruled out: for large
curvatures, the distribution falls off faster than $1/c^2$.  This can
probably be explained by the fact that the eigenvalues and the eigenvectors
are strongly correlated for large disorder in contrast to what was assumed
by Thouless in the derivation of his formula. On the other
hand, Fig.\ref{pofclocLognorm} shows that a 
log--normal distribution does not fit perfectly either. Rather large
deviations are visible for large curvatures. We will address this question
in  more detail in a future work. Nevertheless, we can conclude from
Fig.\ref{pofclocLor} that both $\cb$ and $\cl$ are well--defined quantities
in the localized 
regime \cite{russian}. 

\subsection{Disorder Dependence}

 Before discussing the  disorder
regimes separately, we display in Fig.\ref{FIGcandgofw} an overall plot
of the disorder dependence of  $\cbt$ and $\gl$. Several points can be
observed immediately:\\
First of all, $\cbt$ diverges for small disorder like $1/w^2$. This is a
well-known fact which can be derived from perturbation theory (first Born
approximation \cite{??}). Also, $\cbt$ has the right scaling
behavior of a conductance. In 3D, in the ballistic and diffusive regimes,
$\cbt$ 
increases proportionally to the system size $L$ within the parameter range
provided ($L=6$ to  $L=10$). In the localized regime it
decays with the system size.  
In 3D there is a critical value $w_c\simeq 16.5$ where $\cbt$ becomes
independent 
of the system size, thus indicating the position of the metal--insulator
transition (MIT). Within the error bars it coincides with the well--known
value found by 
Mac\-Kin\-non and Kramer, who examined the scaling behavior of the
transmission 
through disordered samples \cite{macKinnon81}.  We also checked that in 2D
$\cbt$ is 
independent of $L$ in the diffusive and ballistic regimes.

\subsubsection{Ballistic and Diffusive Regimes}
Fig.\ref{FIGcandgofw} shows that $\gl$ obeys the same scaling behavior as
$\cbt$. However, the disorder 
dependence of $\gl$ and $\cbt$  is rather different. Even in the diffusive
regime,   $\gl$ follows
$\cbt$ only over a small disorder interval close to the metal--insulator
transition. The interval's width increases with the system size, but for all
system 
sizes the discrepancy becomes very pronounced in the ballistic
regime, where $\gl$ converges to a
constant value, whereas $\cbt$ keeps diverging.\\

Following our discussion of section \ref{RLT} this result is not
surprising.
 In the ballistic regime, $D$ formally
diverges, as does $\cbt$. Any limitation of the conductance due to the coupling
of the sample to the environment must then result in a deviation from the
conjectured proportionality between $\cbt$   and $\gl$. Clearly, the
discrepancy in the lower disorder limit of the diffusive regime is already
caused by the cross--over to constant $\gl$ due to the  boundary
resistance. \\ 

In order to improve the agreement of $\cbt$ with the conductance, the
latter has to be defined such that it does not incorporate the contact
resistance. We therefore also compared  the disorder dependence of $\gt$
with the one of $\cbt$. As
explained in sec.\ref{RLT},  $\gt$ does not contain the contribution
of the boundary resistance and should be a measure of the  bulk
conductance. It will therefore also diverge when the disorder vanishes.
Whereas it is not clear from the beginning that this divergence will be of
the same kind as the one of $\cbt$, 
Fig.~\ref{FIGcandgtofw} shows that  $\gt$  diverges for small
$w$ indeed with the 
same power as $\cbt$. Both curves follow each other from the diffusive regime
until far into the ballistic regime. \\

In Fig.~\ref{FIGgtofcb} we have
plotted  $\gt$  as function of $\cbt$. The MIT is given in this
plot by the 
point where $4\pi^2\cbt\simeq 4.1$. The points from all sample--sizes 
considered in 3D now fall on one straight line with slope one and this in
the 
diffusive as well as in the ballistic regime \cite{bal}. A fit to a linear
law gives
\begin{equation} \label{eq:fit}
 \gt =(0.99\pm 0.04)\pi\cbt -0.029\pm 0.008
\end{equation}
in remarkable agreement with eq.(\ref{rightp}) \cite{universal}.
The error bars were obtained as  standard deviations from the three
system sizes considered. The conclusion is therefore that $\cbt$ measures
the bulk conductance $\gt$ in both the diffusive and the ballistic regime.

In a recent paper, Casati {\it et al.} \cite{casati96} also study the relation 
between Landauer conductance and curvature distributions for band random matrices
and they find the relation:
\begin{equation}\langle g_L \rangle = (7.5 \pm .4)
{\cal{K}}_{av}\label{brm}\end{equation}
where they define      ${\cal{K}}_{av}$ as the
{\it geometric} average $\exp \langle \ln (|c|/\Delta)\rangle
$\cite{casati96}.
For a distribution of curvatures like \ref{curvature1}, the geometric
average is related to the arithmetic average by\cite{comment}
$${\cal{K}}_{av} =       \exp \langle \ln {|c|\over \Delta}\rangle
= {1 \over 2} \langle { |c|\over \Delta}\rangle$$
Using the relation \ref{g4} we get the following result:
$$\langle g_L \rangle= 2 \pi {\cal{K}}_{av}$$
This factor $2 \pi$ satisfactorily explains
the numerical result \{eq. \ref{brm}\} found by these authors.

\subsubsection{Localized regime}
Fig.\ref{FIGgtofcb} shows that the power between $\cbt$ and $\gt$ changes
at the MIT. 
We obtain approximately $\gl\propto\cbt^{1.2}$.  
However,  in the localized regime $\cbt$ and $\gt$
might not be the right quantities to look at. At least from the conductance
it is known that in this regime the function with the right scaling behavior
is $\langle \ln g_L\rangle$ \cite{irrelev}, not $\gl$. Since the favored
log--normal distribution of curvatures is due to the same reason as the
log--normal distribution of the conductances, namely the exponentially
decaying wave functions with normally distributed localization length, one
might suspect that a similar statement holds 
for the curvatures as well. We therefore also examined $\langle \ln
g_L\rangle$ 
and $\cl$ as functions of disorder and system size. Fig.\ref{clandgl} shows
the result: first of all both  quantities are 
proportional to the system size, and secondly proportional to each other.
Plotting $\langle \ln g_L\rangle$ versus $\cl$ (see Fig.\ref{lngvslnc}) yields a 
straight line that is best
approximated by the linear law
\begin{equation} \label{proploc}
\langle \ln g_L\rangle\simeq 1.7 \cl-2.5\,.
\end{equation} 
Again, the validity of this equation extends over several orders of
magnitude of the conductance. However, the prefactor seems to decrease
slightly but systematically with the system size. A  law like \ref{proploc}
was also
reported in \cite{casati96} with a similar prefactor (1.73) for banded random
matrices. Nevertheless, Ref.\cite{casati96} also reports a prefactor 2.0 for
an Anderson model. We do not have any explanation for
this difference besides the fact that in contrast to our box--distributed
disorder  the disorder was
Gaussian distributed in \cite{casati96}.
The found behavior is a priori surprising. Assuming that in the localized
regime the flux dependence of each energy level is purely sinusoidal
\cite{sinus} one deduces that $\langle \overline{i^2(\varphi)}\rangle\propto
\langle c^2\rangle$, where $i_\alpha(\varphi)=-\partial
e_\alpha/\partial\varphi$ and the overline indicates a flux average. Since
$\langle \overline{i^2(\varphi)}\rangle$ can be related to the Kubo
conductance, one would expect a quadratic relation $\gl\propto \langle
c^2\rangle$\cite{akkermans94}.
 
\section{Conclusion}\label{concl}
We have examined numerically the relation between  level
curvatures  and
conductances for disordered systems. We showed that  in the diffusive
regime a proportionality between the dimensionless mean absolute curvature
$\cbt$ and the average Landauer--B\"uttiker 
conductance  $\gl$ holds if the system is large enough so that  the influence
of the boundary resistance can be neglected. In the ballistic regime, the
boundary resistance can never be neglected and leads to a strong violation
of the proportionality. In the limit of zero disorder it completely
dominates the total resistance and limits  $\gl$ to the number of open
channels, whereas $\cbt$ diverges in the 
same limit. However, for all system sizes a proportionality between a
properly defined bulk conductance $\gt$ and $\cbt$ could be established that
holds 
in the same form in the diffusive and ballistic regimes. This shows that in
these regimes level
curvatures measure  a conductance that is entirely due to
the dynamics in the bulk of the sample and therefore not influenced by
details of the
measurement setup, like the number of leads and the way they are attached. 
In the localized regime, we found a proportionality between $\cl$ and
$\langle \ln g_L\rangle$.

{\it Acknowledgements: }
We would like to thank M.B\"uttiker for useful discussions and
correspondence. D.B. is grateful to A.MacKinnon and E.Hofstetter for
hospitality during his stay in London, where this work was initiated. This
work was partially supported by the French Academy of Science.

\begin{figure}
\protect\caption{\label{pofcbal} The curvature distribution in the ballistic
regime. The full line is the prediction of eq.(\protect\ref{curvature1})
known to be valid in the diffusive regime.
Deviations at small curvature are visible (see inset). 
$6\times 6\times 6$ sites, $w=1.0$, 4000 disorder realizations.} 
\end{figure}

\begin{figure}
\protect\caption{\label{pofclocLor} The curvature distribution in the
localized regime with a fit to a Cauchy distribution (dashed line). This
plot shows that 
$P(c)$ decays faster than $1/c^2$ for large curvatures. ($6\times 6\times 6$,
$w=50$, 4000 disorder realizations).}
\end{figure}

\begin{figure}
\protect\caption{ Distribution of $\ln |c|$ in the
localized regime with a fit to a Gaussian distribution (corresponding to a
log--normal distribution for $|c|$). Same parameters as in
Fig.\protect\ref{pofclocLor}\label{pofclocLognorm}.} 
\end{figure}

\begin{figure}

\protect\caption{The overall disorder dependence of $\langle
|\tilde{c}_\alpha|\rangle $ (top) 
  and 
  $\langle g_L\rangle$ (bottom) for different system sizes: diamonds
$6\times 6\times 6$, 
  circles $8\times 8\times 8$, and triangles $10\times 10\times 10$. The
  straight lines of $\langle |\tilde{c}_\alpha(w)|\rangle $ in the
logarithmic 
plot correspond to 
  a 
  $1/w^{2}$ divergence for small $w$. Full lines are guides to the eye
only. For clarity the $\langle
|\tilde{c}_\alpha(w)|\rangle $ curves were shifted by an arbitrary factor
$4\pi^2$.\label{FIGcandgofw}}  
\end{figure}

\begin{figure}

\protect\caption{\label{FIGcandgtofw} The overall disorder dependence of 
$\langle
|\tilde{c}_\alpha|\rangle $ (top) and 
  $\gt $ (bottom) for different system sizes (same symbols as in
  FIG.\protect\ref{FIGcandgofw}). $\langle\tilde{g}_L(w)\rangle $ diverges in
the same manner as 
$\langle |c_\alpha(w)|\rangle $ for small $w$.  Full lines are guides to the
eye 
only.} 
\end{figure}

\begin{figure}

\protect\caption{\label{FIGgtofcb} The conductance $\langle  \tilde{g}\rangle$ 
plotted against
$\langle |\tilde{c}_\alpha|\rangle $ for different system sizes (same symbols as
in Fig.
\protect\ref{FIGcandgofw}). The diffusive regime starts with the critical
  mean curvature $4\pi^2\cbt \simeq 4.1$. In this regime
and the ballistic 
  regime the dependence is very well fitted by the same linear law
  $ \gt=(0.99\pm 0.04)\pi\cbt -0.029\pm 0.008$ (full line).}
\end{figure}

\begin{figure}
\protect\caption{\label{clandgl} $\cl$ (top) and $\langle \ln g_L\rangle$
(bottom) as a
function of 
disorder 
for three different system sizes in the localized regime (same symbols as
in Fig.
\protect\ref{FIGcandgofw}).} 
\end{figure}

\begin{figure}
\protect\caption{\label{lngvslnc} $\langle\ln g_L\rangle$ as a function of
$\cl$ in the 
localized regime for three different system sizes (same symbols as
in Fig.
\protect\ref{FIGcandgofw}).}  
\end{figure}

\begin{references}
\bibitem{RMT} In the diffusive regime described by the Random Matrix Theory,
this is implied by the statistical independance of energy levels and
energy eigenstates.
\bibitem{remark1} In the literature, it is also sometimes defined as $g_K =
\hbar G_K/e^2$. The notation we chose here is consistent with the
dimensionless Landauer--B\"uttiker conductance defined below, eq.(\ref{Gl1}).
\bibitem{bal} The ballistic regime can be effectively determined from
Fig.\ref{FIGcandgofw} as the regime where the conductance starts to be
limited by the inverse contact resistance.
\bibitem{comment} The distribution of conductances is gaussian: the
arithmetic and the geometric averages are identical as long as they are
large compared to the width of the distribution.
\bibitem{universal} We draw attention to the fact that the prefactor $\pi$
is not universal but depends on the sample geometry \cite{fyodorov95}. In
our calculations all samples had the same ring geometry.
\bibitem{irrelev} In the localized regime it is irrelevant whether or not the
boundary resistance is included in the conductance: $\gl=\gt$ up to
exponentially small corrections.
\bibitem{sinus} This seems very natural since in the diffusive regime the
harmonics decrease exponentially as $\exp(-m\sqrt{\Delta/E_c})$.
\end{references}
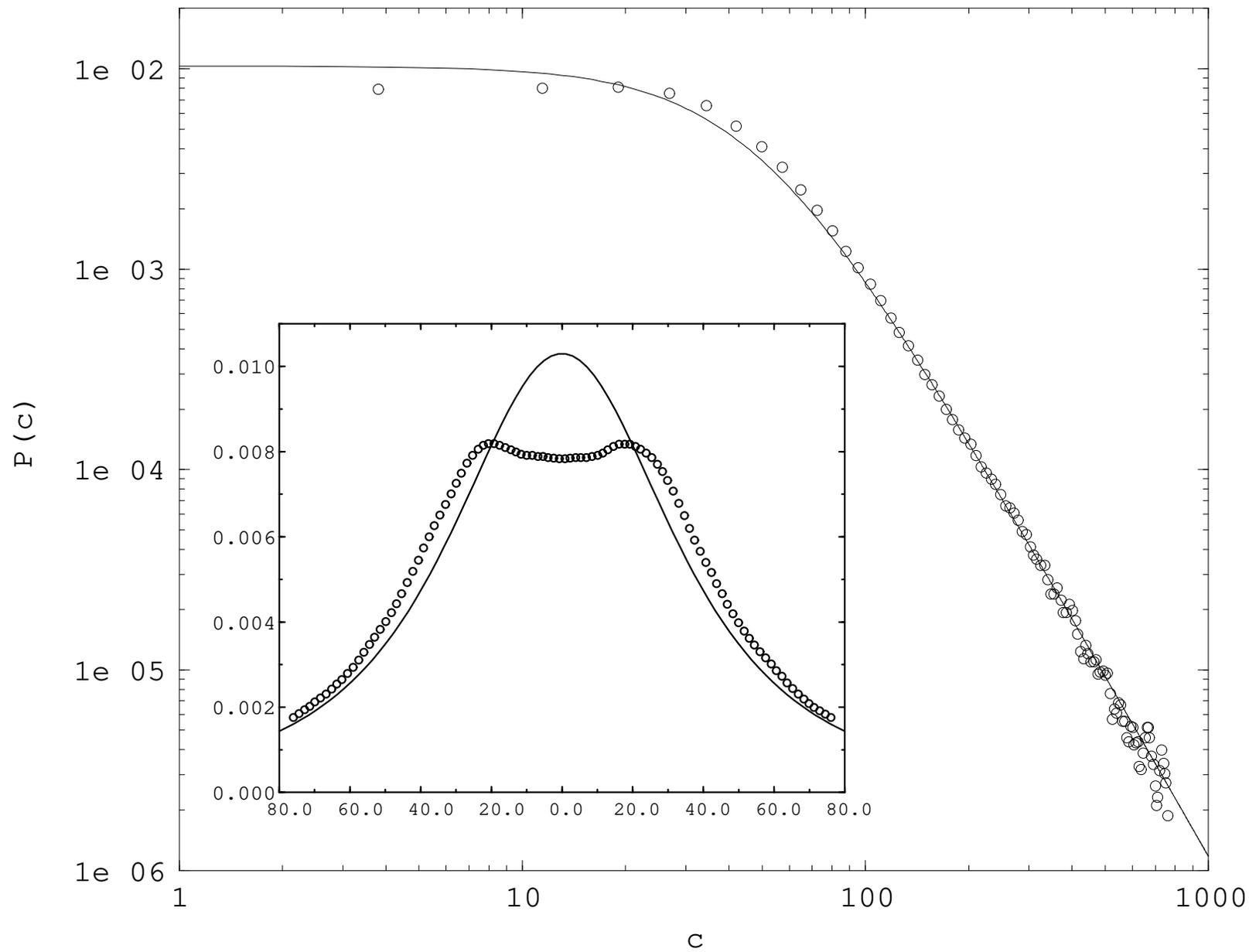
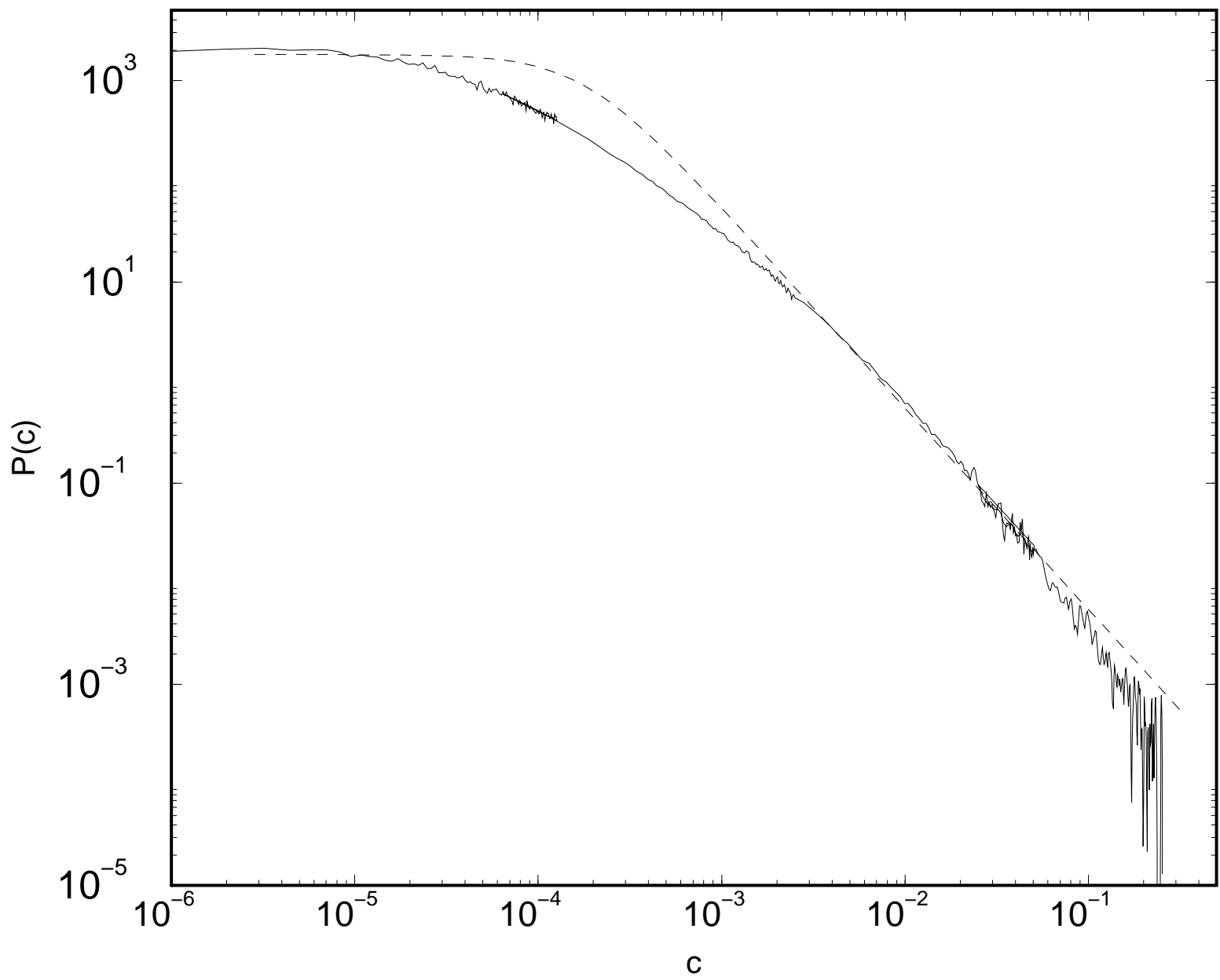
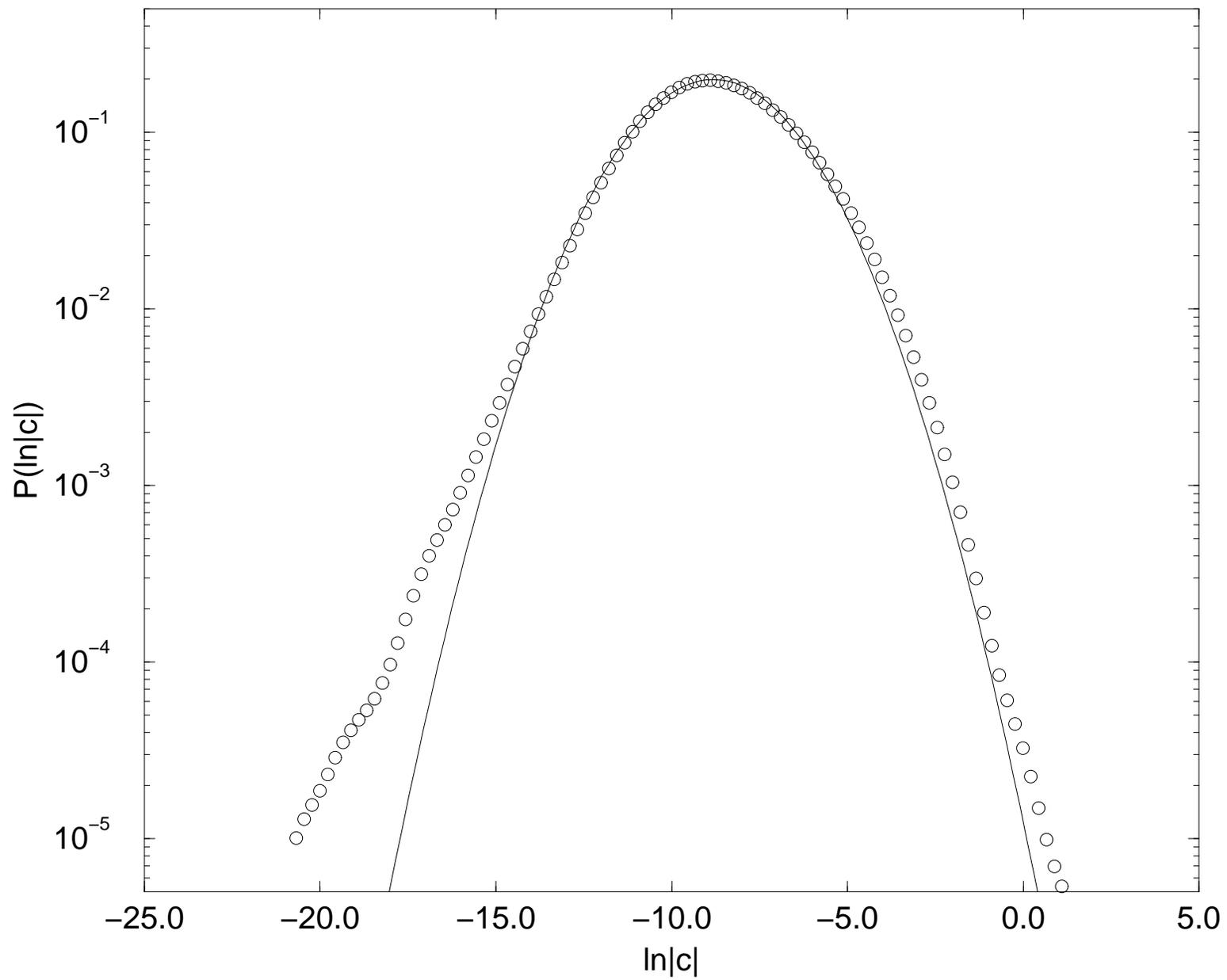
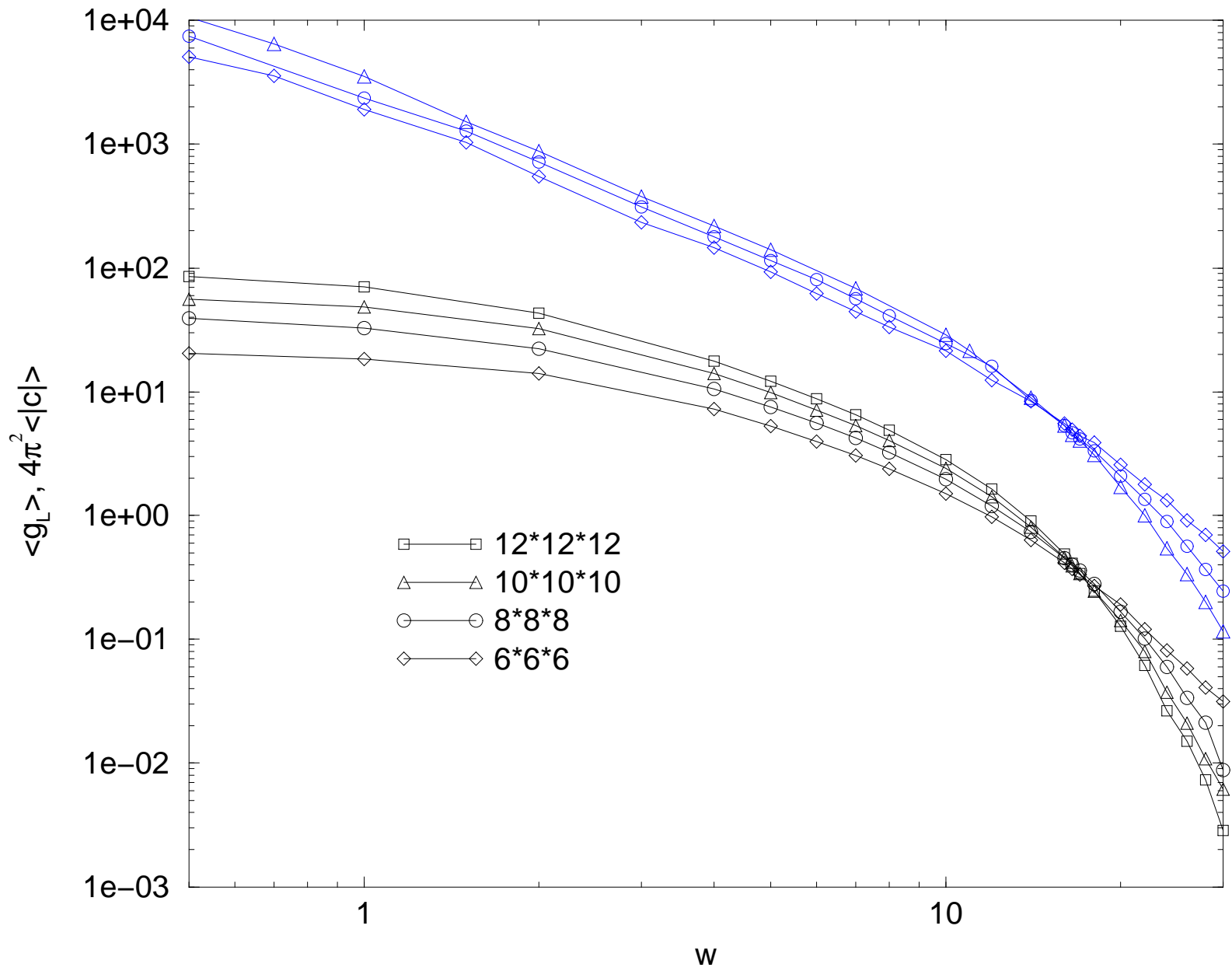
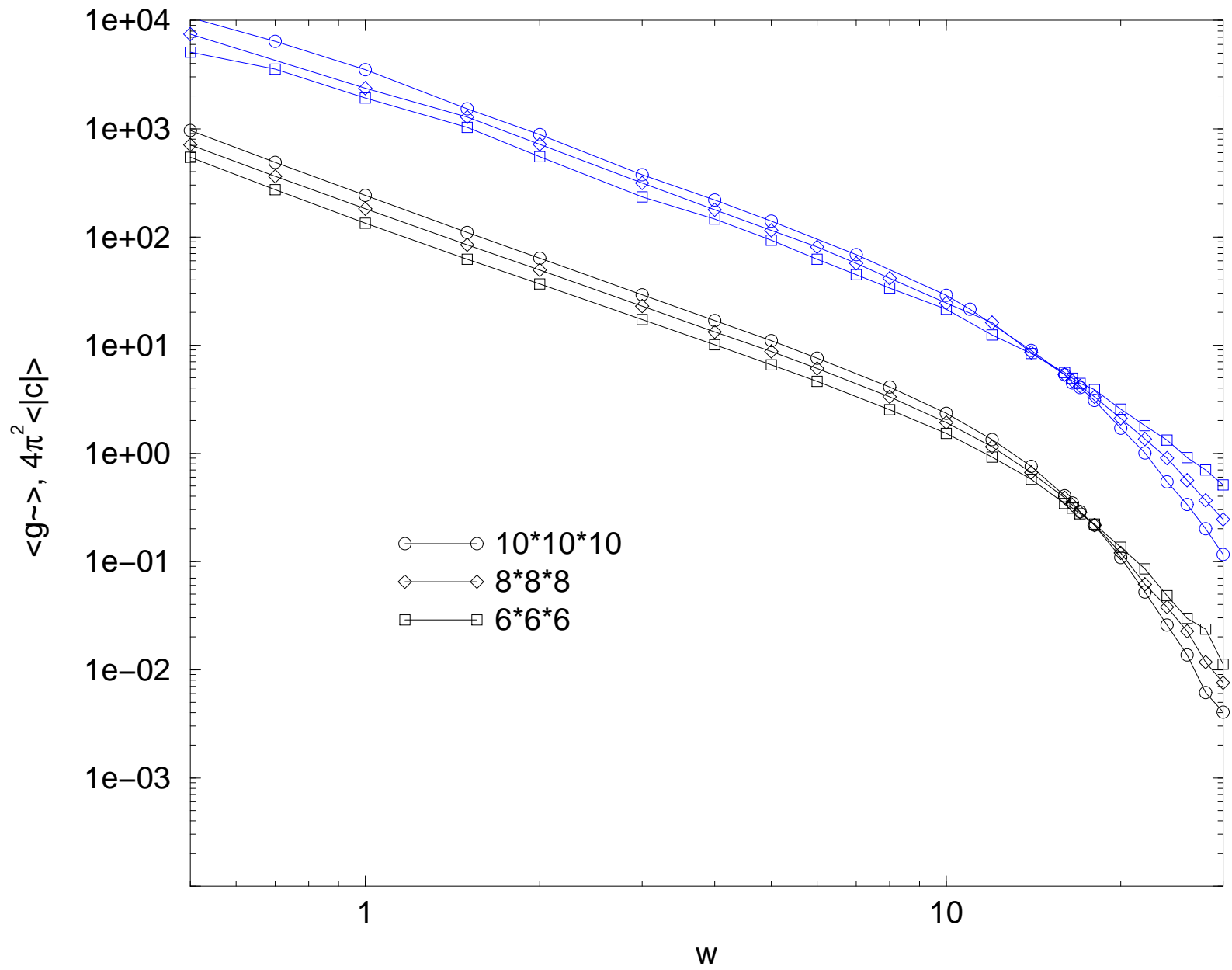
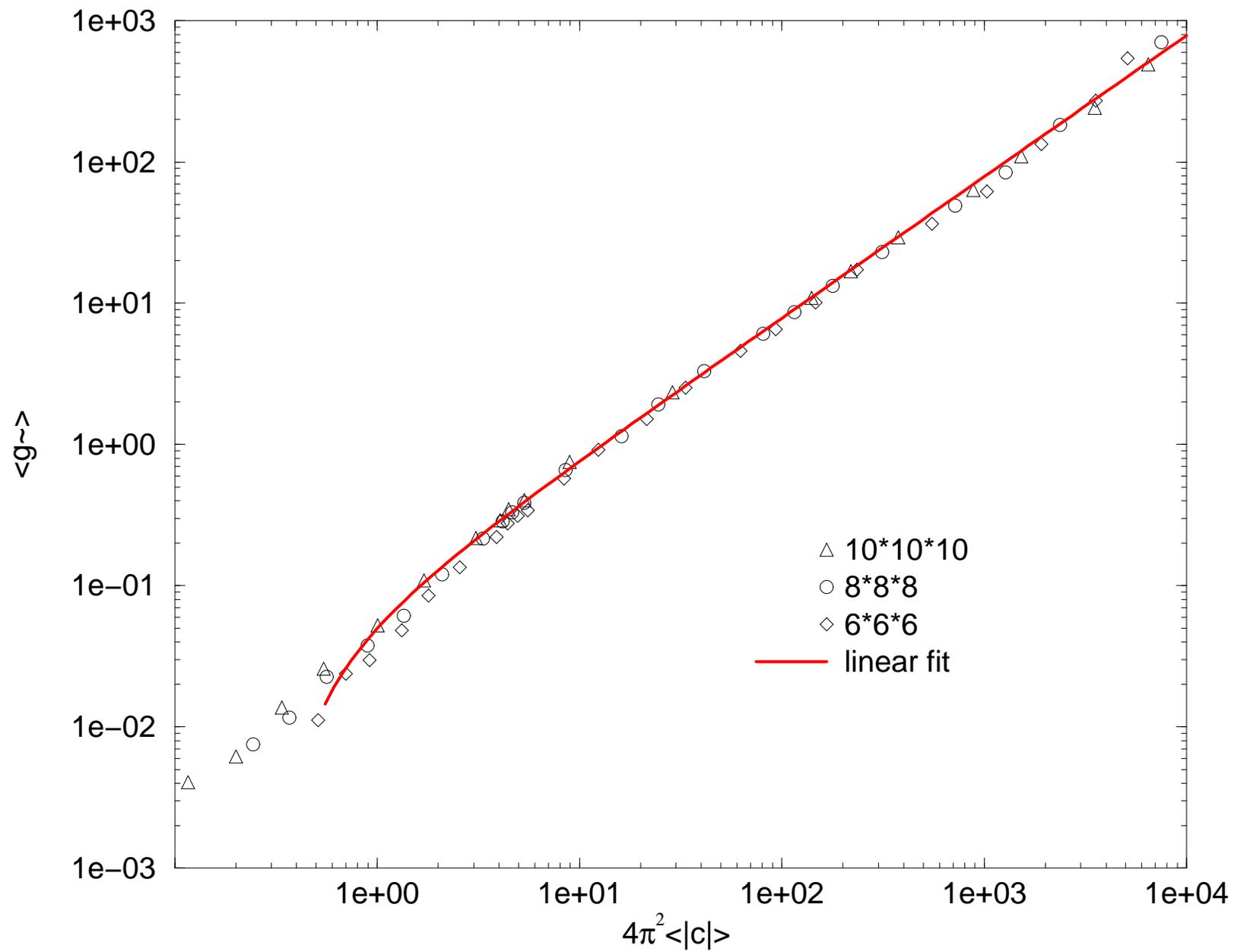
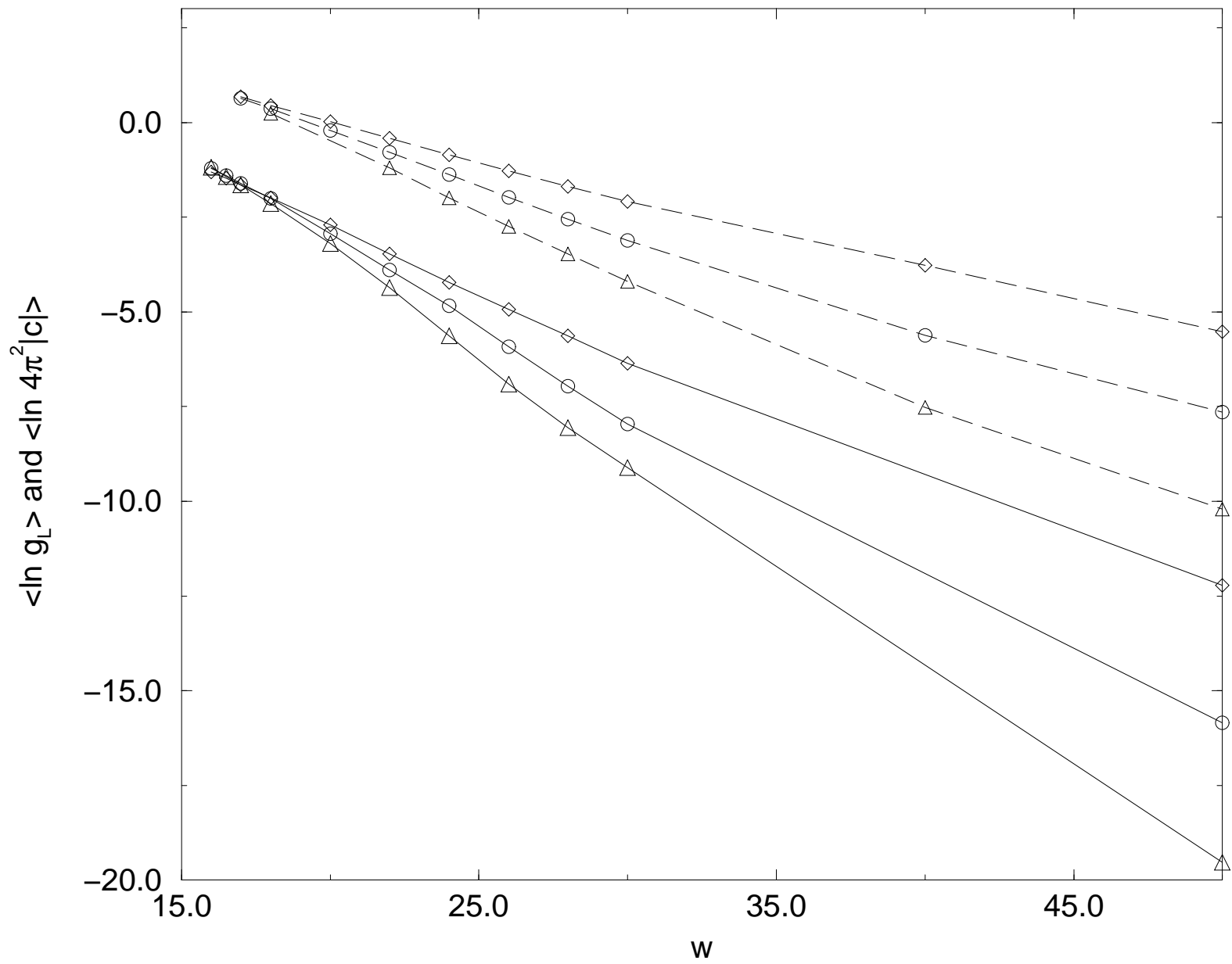
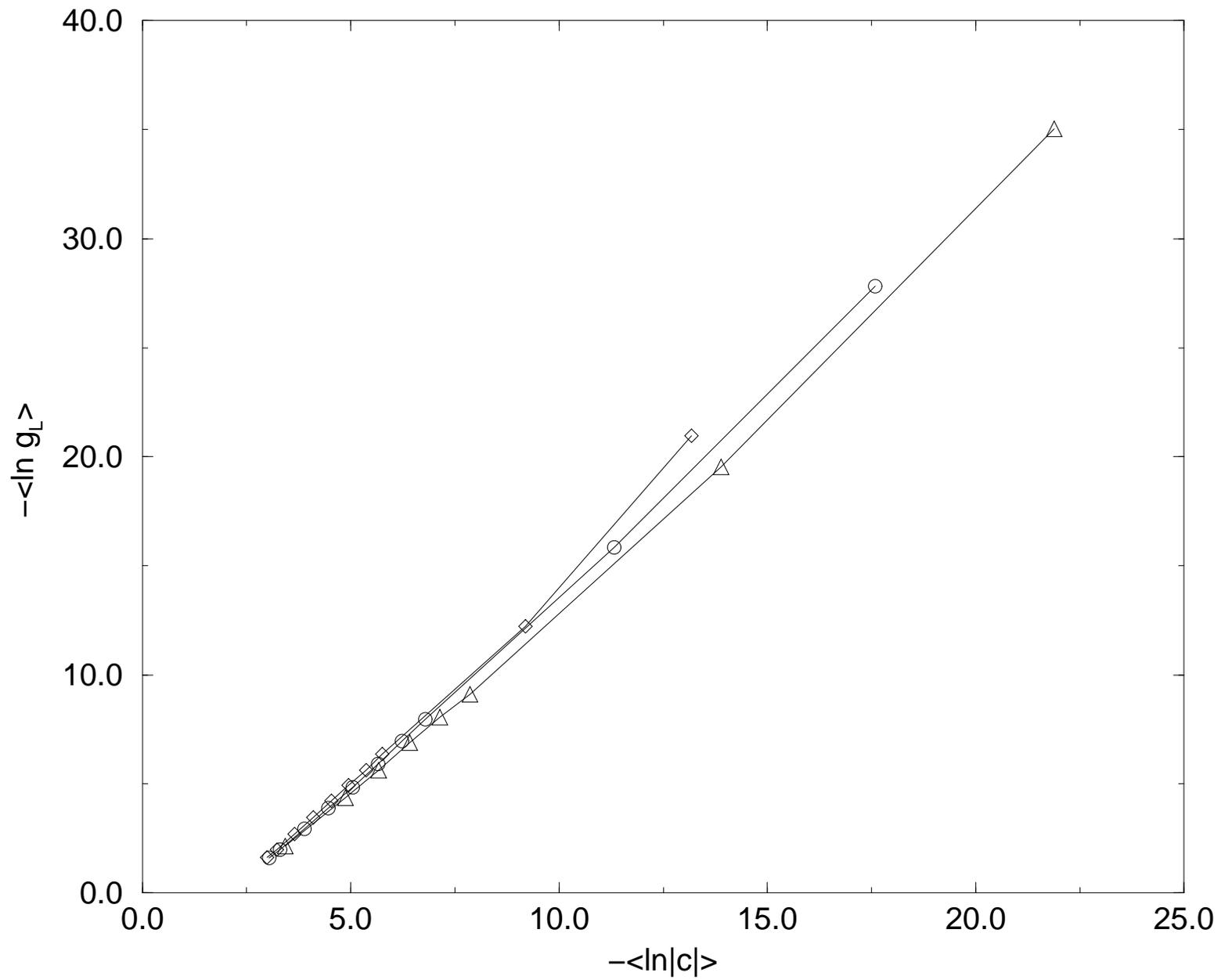
\end{document}